\documentclass{natureHack}

\usepackage{authblk}
\usepackage{amsmath}
\usepackage{amssymb}
\usepackage{hyperref}
\usepackage{capt-of}
\usepackage{multirow,booktabs}
\usepackage{url}
\usepackage{multibib}
\usepackage{graphicx}

\newcites{M}{Methods Bibliography}

\title{All planetesimals born near the Kuiper Belt formed as binaries}

\author{Wesley C. Fraser$^{1}$, 
Michele T. Bannister$^{1}$, 
Rosemary E. Pike$^{2}$, \\
Michael Marsset$^{1,3}$,
Megan E. Schwamb$^{4}$, 
JJ Kavelaars$^{5}$, \\
Pedro Lacerda$^{1}$,
David Nesvorny$^{6}$,
Kathryn Volk$^{7}$, \\
Audrey Delsanti$^{3}$, 
Susan Benecchi$^{8}$, 
Matthew J. Lehner$^{2}$, 
Keith Noll$^{9}$, 
Brett Gladman$^{10}$, \\
Jean-Marc Petit$^{11}$, 
Stephen Gwyn$^{12}$, 
Ying-Tung Chen$^{2}$, \\
Shiang-Yu Wang$^{2}$, 
Mike Alexandersen$^{2}$, 
Todd Burdullis$^{13}$, \\
Scott Sheppard$^{14}$, 
Chad Trujillo$^{15}$}

\begin{document}

\maketitle

\begin{affiliations}
\item Queen's University Belfast
\item Institute of Astronomy and Astrophysics, Academia Sinica
\item Aix Marseille Universit\'e, CNRS, LAM, Laboratoire d'Astrophysique de Marseille
\item Gemini Observatory, Northern Operations Center
\item University of Victoria
\item Department of Space Sciences, Southwest Research Institute
\item University of Arizona
\item Planetary Science Institute
\item NASA Goddard Space Flight Center
\item University of British Columbia
\item Institut UTINAM, UMR6213 - CNRS - Universit\'e Bourgogne Franche Comt\'e, F-25000 Besan\c{c}on, France
\item National Research Council of Canada
\item CFHT Canada France Hawaii Telescope
\item Carnegie Institution for Science
\item Northern Arizona University

\end{affiliations}

\begin{abstract}
The cold classical Kuiper Belt Objects have low inclinations and eccentricities\cite{Brown_2001,Kavelaars_2009}, and are the only Kuiper Belt population suspected to have formed in-situ\cite{Parker_2010}. Compared to the dynamically excited populations which exhibit a broad range of colours, and a low $\sim 10\%$ binary fraction\cite{2008ssbn.book..345N}, cold classical objects typically possess red optical colours\cite{Gulbis_2006}, with $\sim30\%$ of the population  found in binary pairs\cite{Grundy_2011}; the origin of these differences remains unclear\cite{Benecchi_2009,Fraser_2012}. We report the detection of a population of blue coloured, tenuously-bound binaries residing amongst the cold classical objects. Here we show that widely separated binaries can survive push-out into the cold classical region during the early phases of Neptune's migration\cite{Nesvorn__2015}. The blue binaries may be contaminants, originating at $\sim38$~AU, and could provide a unique probe of the formative conditions in a region now nearly devoid of objects. The idea that the blue objects, which are predominantly binary, are products of push-out requires that planetesimals form entirely as multiples. Plausible formation routes include planetesimal formation via pebble accretion\cite{Shannon_2016} and subsequent binary production through dynamical friction\cite{Goldreich_2002}, and binary formation during the collapse of a cloud of solids\cite{Nesvorn__2010}.
\end{abstract}

We report new colour measurements of 16 cold classical Kuiper Belt Objects (CCKBOs; see methods for our definition of a CCKBO), including the colours of 4 wide binary systems detected by the Outer Solar System Origins Survey\cite{Bannister_2016} (OSSOS) and the Colours of OSSOS Gemini Program (see Fig. 1, and Supplementary Table 1). While the known single or unresolved binary (hereafter both referred to as singles) CCKBOs exhibit predominantly red surface colours with spectral slopes $s>15\%$, the known CCKBO  binaries exhibit a different colour distribution, extending to  $s\sim0$\%. The 5 bluest CCKBOs are all resolved binary pairs. By contrast, only one single object has  a spectral slope  $s<17\%$, the slope at which the full Kuiper Belt colour distribution bifurcates into two classes, historically labelled as \emph{red} and \emph{blue}\cite{Peixinho_2012,Fraser_2012}.  Scaling from the binary sample, 12 blue single CCKBOs are expected. Monte Carlo simulations demonstrate that the probability that the binaries and singles share the same colour distribution is only 0.1\% (see Methods).

The blue binary CCKBOs are found throughout the cold classical region, and exhibit a heliocentric orbit distribution that is similar to all other CCKBOs (see Fig. 2). Unlike the red binary CCKBOs which exhibit a broad range of binary orbital separations from $4\times10^4$~km down to the resolvable limit, the blue binaries are entirely devoid of closely bound pairs with $a_{\textrm{bin}}<5500$~km (see Fig. 1). That blue binaries are only found on widely separated orbits is unlikely to be an observational bias. At least half of the objects identified as binary have been observed with the Hubble Space Telescope which is sensitive to binary separations of $\gtrsim2000$~km at cold classical distances. If the blue binaries shared the same binary semi-major axis distribution as the red binaries, the probability that all detected blue binaries have $a_{\textrm{bin}}>5500$~km is only 2\%. While it seems unlikely that the source of their colours and separations are directly linked, the preference of blue CCKBOs to be binary with large $a_{\textrm{bin}}$ is unquestionable.

The paucity of widely separated binaries and the low binary fraction of the dynamically excited population means that the observed blue CCKBOs, which are found almost entirely as binaries, are not contamination from the low-inclination tail of the current excited population\cite{Brown_2001}. The small KBOs which possess higher inclinations and eccentricities than the CCKBOs occupy a broad range of colours, $0\lesssim s \lesssim50$~\% and have a low $\sim10\%$ binary fraction\cite{2008ssbn.book..345N}. Rather, the blue binaries must have a different formative, or evolutionary history than the excited objects.

Collisional dredge up of fresh, blue coloured icy material is one possible mechanism for the conversion of a red CCKBO into a blue  CCKBO. This mechanism is largely excluded, as all observed KBO binaries, including those we report here, exhibit equally coloured binary components\cite{Benecchi_2009}, requiring that any colour altering collisions must {\it always} alter both bodies simultaneously. Ejecta simulations demonstrate that significantly more of the ejected material is re-accreted by the impacted body than is by the secondary (see Methods). Therefore, collisions by other KBOs can not cause significant alteration of an object's colour while still preserving the colour equality of the binary components\cite{Benecchi_2009}. 

A possible alternative source of impactors are collisions between bound members of higher multiplicity systems, like the triple system, 1999 TC36\cite{Benecchi_2010}. Higher multiplicity systems are a natural by-product of planetesimal formation models in which large planetesimals form during the gravitational collapse of a bound cloud of solid material\cite{Nesvorn__2010}. When the collapse results in binary systems like those observed in the classical region, both binary members experience impact velocities as high as $\sim50 \mbox{ m s$^{-1}$}$ from the remaining cloud debris (see Methods). Even in this scenario, impacts by small bodies are always more frequent than by larger bodies, and so the binary colour symmetry would be broken.

The known members of the Haumea collisional family are water-ice rich KBOs that appear to be the remnants of a larger disrupted parent object\cite{Brown_2007}. All family members exhibit similarly blue optical surface colours as do the blue CCKBO binaries, suggesting the possibility that the blue binaries are the fragments from the disruption of a larger KBO. Simulations of disruptive collisions produce binary systems from the ejecta fragments\cite{Durda_2004}. Those simulations however, mainly produce single objects, consistent with the observation that all small Haumea family members are single objects. Given the dearth of single blue CCKBOs, it seems unlikely that the blue CCKBO binaries are the fragments of a large disrupted object. 

Recent simulations have demonstrated that the heliocentric orbital structure of the cold classical region including the so-called kernel\cite{Petit_2011}, or concentration of objects at $\sim44.5$~AU, can be accounted for if Neptune's outward migration were predominantly smooth, but disrupted by a large jump of $\sim0.5$~AU\cite{Nesvorn__2015}. During this migration, a small fraction of objects can be swept up in the 2:1 mean-motion resonance (MMR) and be transported outwards, avoiding any close encounters with Neptune which would otherwise disrupt the binary pairs\cite{Parker_2010}.  Our n-body simulations demonstrate that binaries as widely separated as the most tenuously bound known binary, 2001 QW322\cite{Petit_2008}, are robust against being unbound during this push-out process; binaries with initial binary semi-major axis to mutual Hill sphere ratios, $a_{\textrm{B}}/R_{\textrm{H}} \leq 0.25$, survived being pushed out into the cold classical region (see Fig. 3). The details of our simulations are presented in the Methods.

The idea that the blue CCKBO binaries are a contaminant that was pushed out during Neptune's migration on to cold classical orbits  agrees well with the hypothesis that the blue-red bifurcation of the excited KBO colour distribution\cite{Peixinho_2012,Fraser_2012} is a result of an object's heliocentric formation distance\cite{Brown_2011}. In our simulations, surviving binaries were pushed out by no more than 6~AU, originating in the $\sim38-40$~AU range. Similar simulations suggest that the red coloured, widely separated binary, 2007 TY430, which now resides in the 3:2 MMR plausibly originated at $\sim37-39$~AU but could have originated even further out\cite{Nesvorn__2015}. This would require that the distance inside which blue binaries originated was only a handful of AU inside the current inner edge of the cold classical region. Literature estimates\cite{Brown_2011} of the distance outside of which small objects are cold enough to retain NH$_3$ against sublimation loss is $\sim34$~AU suggesting that early sublimation loss of ammonia may play a role in the production of the different surfaces of these two populations.

The push-out scenario we describe would have the startling implication that virtually \emph{all} planetesimals that formed in the region from which the blue binaries originated must either have formed as binaries or higher multiplicity systems, or attained high multiplicity before the push-out occurred. This is required by the fact that all but one of the blue CCKBOs are found in binary pairs. Our scenario makes a number of predictions. Firstly, there should be a concentration of blue binaries alongside the kernel\cite{Petit_2011} which appears to be populated by the same push-out process\cite{Nesvorn__2015}. Secondly, there should be a paucity of blue binaries beyond the kernel's outer edge at $\sim45$~AU, a prediction that is consistent with the known objects. Finally, compared to the population interior to the kernel, a high blue:red fraction for widely separated binaries on low eccentricity and inclination orbits in the 2:1 MMR would disfavour our push-out scenario, as the low excitation 2:1 resonators should be predominantly populated from objects that resided at and beyond the kernel distance after the jump occurred.

Recently, the idea of producing large planetesimals from the collapse of gravitationally bound clouds of solids has been put forth as a potential solution to avoiding a number  of barriers that hinder planetesimal growth\cite{Johansen_2007}. Formed through instabilities driven by the interaction of gas and solids in a disk, the subsequent collapse of these clouds spontaneously produces a bound system of two or more objects\cite{Nesvorn__2010}. This presents a natural solution to forming a planetesimal population with a high binary fraction, though this mechanism may be unable to produce the large population of retrograde KBO binaries\cite{Schlichting_2008b,Nesvorn__2010}. If KBOs formed in this way, there will be relatively many more single objects with radii smaller than the individual primary components, as, during the collapse process, a large number of single objects with radii $\sim10-50\%$ that of the primary body are ejected from the collapsing system. These objects would have brightnesses of $\gtrsim 25$~magnitude, fainter than any colour or binary survey to date.

An alternative scenario for the creation of a nearly 100\% binary fraction is single planetesimal growth through pebble accretion with subsequent binary production via the L2s processes, or dynamical friction from the residual pebble population\cite{Goldreich_2002}. Pebble accretion simulations can reproduce the CCKBO size distribution\cite{Shannon_2016,Fraser_2014} while maintaining a sub-hill velocity dispersion and a surface density of small planetesimals throughout the majority of the planetesimal growth phase. These conditions are conducive to binary formation through dynamical friction; rough estimates suggest that in these conditions, the fractional binary formation rate is on the order of $\sim10^{-6}$ per year, implying a virtual 100\% binary fraction after just one million years.

If the blue binaries are contaminants of a push-out process, then those systems are easily identifiable probes of the early disk conditions interior to the cold classical range. Our simulations demonstrate that tightly bound binaries survive cold classical implantation more frequently than do initially wider separated pairs. During the push-out process, excitation in binary eccentricity occurred for objects with $a_{\textrm{bin}}\gtrsim0.1R_{\textrm{H}}$, with only modest change in binary semi-major axis. With the exception of binary systems with the largest eccentricities, the typical dynamical evolution of a binary under the effect of tides and collisions results in no significant change in binary semi-major axis. Therefore, the push-out scenario would also demand that the region from which blue binaries originated must also be entirely devoid of closely separated binaries. This would imply the existence of a gradient in the formative conditions between $\sim 37$ and $\sim 44$~AU. If KBOs were produced by cloud collapse, it may be that the formative conditions interior to the cold classical region resulted in clouds with higher angular momenta than formed further out, which on average would produce more widely separated binaries. Such a scenario could be explained if the cloud masses decreased with heliocentric distance from $37$ to $44$ AU (see Methods). In the L2s binary formation mechanism, the orbital energy loss due to dynamical friction is a strong function of the pebble disk surface density. It may be then that conditions at $\sim37$~AU were just right to result in a near 100\% binary fraction, but insufficient to produce a large population of binaries with $a_{\textrm{bin}}\lesssim0.05R_{\textrm{H}}$.

\newpage

\textbf{References}
\newcounter{mainbib}


\textbf{Addendum}
 \textbf{Acknowledgements} This is work is based in part on observations from the Large and Long Program GN-2014B-LP-1, obtained at the Gemini Observatory, which is operated by the Association of Universities for Research in Astronomy, Inc., under a cooperative agreement with the NSF on behalf of the Gemini partnership: the National Science Foundation (United States), the National Research Council (Canada), CONICYT (Chile), Ministerio de Ciencia, Tecnolog\'{i}a e Innovaci\'{o}n Productiva (Argentina), and Minist\'{e}rio da Ci\^{e}ncia, Tecnologia e Inova\c{c}\~{a}o (Brazil). This work is also based on observations obtained  with  MegaPrime/MegaCam,  a  joint  project  of  the Canada-France-Hawaii  Telescope (CFHT) and CEA/DAPNIA, at CFHT which is operated by the National Research Council (NRC) of Canada, the Institute National des Sciences de l'Universe of the Centre National de la Recherche Scienti que (CNRS) of France, and the University of Hawaii. A portion of the access to the CFHT was made possible by the Institute of Astronomy and Astrophysics, Academia Sinica, Taiwan. This research used the facilities of the Canadian Astronomy Data Centre operated by the National Research Council of Canada with the support of the Canadian Space Agency.  This paper includes data gathered with the 6.5 meter Magellan Telescopes located at Las Campanas Observatory, Chile.
 
The authors wish to recognize and acknowledge the very significant cultural role and reverence that the summit of Mauna Kea has always had within the indigenous Hawaiian community. We are most fortunate to have the opportunity to conduct observations from this mountain.

MES was supported by Gemini Observatory, which is operated by the Association of Universities for Research in Astronomy, Inc., on behalf of the international Gemini partnership of Argentina, Brazil, Canada, Chile, and the United States of America.  MES was also supported in part by an Academia Sinica Postdoctoral Fellowship.

This research made use of the Giorgini, JD and JPL Solar System Dynamics Group, NASA/JPL Horizons On-Line Ephemeris System, \url{http://ssd.jpl.nasa.gov/?horizons}.

We thank Jaime Coffey for her assistance in acquiring images of 2006 BR284.

 \textbf{Competing Interests} The authors declare that they have no competing financial interests.
\textbf{Correspondence} Correspondence and requests for materials should be addressed to W. C. Fraser.~(email: wes.fraser@qub.ac.uk).

\textbf{Author Contributions Statement}
T. B., S. S. and C. T. acquired telescope observations enabling the binarity of 2002 VD131 to be detected. B. G., J. K., J.-M. P., and S. G., designed and with help from M. T. B., S.-Y. W, M. A., and Y.-T. C., operated the OSSOS survey, the detections of which were critical to the success of the Col-OSSOS survey.  K. V. provided dynamical analysis of all newly reported objects in this manuscript. K. N. provided confirmation of the statistical methods used in this paper. A. D., S. B., M. J. L., helped create the initial design and science plan for the Col-OSSOS survey. D. N. provided numerical simulations required to understand the formation of binary planetesimals. M. T. B., R. E. P., M. M., M. E. S.,  and J. K. were instrumental in the design and operations of the Col-OSSOS survey, from which the majority of colour measurements we report, were acquired. P. L. wrote some sections of the methods, and provided insight into understanding our observations within the cloud collapse model. W. C. F. is the PI of the Col-OSSOS survey, and with the help of R. E. P. and M. B., produced all photometry reported in this manuscript. W. C. F. also produced the numerical migration, and ejecta transfer simulations reported here, and wrote the majority of the paper.

\textbf{Data Availability Statement}
Raw data were generated at the Canada-France-Hawaii, Gemini-North, and Magellan Telescopes. Raw telescope imagery from the Gemini-North and Canada-France-Hawaii telescopes, and all associated calibration files are publicly available at the Gemini archive \\(\url https://archive.gemini.edu/searchform), and the Canadian Astronomy Data Centre (\url http://www.cadc-ccda.hia-iha.nrc-cnrc.gc.ca/en/), respectively. Processed imagery and calibration data supporting the findings of this study are available from the corresponding  author (W. C. F.) upon request.


\begin{methods}
\subsection{Colours and  Binary Properties}
Here we consider barycentric orbital elements extracted from the JPL Horizons Web-service\footnote{\url{http://ssd.jpl.nasa.gov/?horizons}} on July 1st, 2016 (see Fig. 2). We only consider those objects with arcs long enough for an eccentricity to be reported. The exact definition of a cold classical has not been agreed upon  (compare \citeM{Elliot_2005} and \citeM{Batygin_2011} for example). The dynamically fragile, widely separated binaries however, are commonly associated with the cold populations. Therefore, we adopt a definition for cold classical object similar to that commonly used in the literature \citeM{Elliot_2005} that encompasses all known binary objects near the cold belt. Specifically, we define a CCKBO as any object not in a mean-motion resonance with Neptune, possessing a semi-major axis $42\leq a \leq 47.5$~AU, a perihelion distance $q>36$~AU, and an ecliptic inclination $i<6^\circ$; resonant behavior of new and known objects was searched for as in \citeM{Gladman_2008}. Critically, we emphasize that small variations in our definition of cold classical do not appreciably alter the significance of the results presented in this paper. 

For our sample of colours, we consider newly reported observations as well as previously reported colours. Reported KBO colours were selected from the literature based on their reliability. We adopt optical colour measurements from the Hubble Space Telescope (HST) reported by \citeM{Benecchi_2009,Benecchi_2011,Fraser_2015} and only those measured in a single orbit. We also consider the colours reported by \cite{Peixinho_2015}, and the mean (g'-r') colours of 2001 QW322 reported in \citeM{Petit_2008}. Finally, we include the measurement of cold classicals 1999 HS11, 2001 KK76, and 2002 VD131 reported by \citeM{Gulbis_2006}. These datasets have all measurements of a target taken with a short time span ($<1.5$~hours for ground based observations; within a $<44$~minute observability window for the HST observations) and are least likely to suffer from any rotational variations a target may exhibit. Only colours reported in the $\sim500-850$~nm range were considered. Specifically, we consider Johnson-Cousins (V-R) and (V-I), Sloan (g'-r'), and HST (F606w-F814w) colours. Spectral slopes and uncertainties were calculated using the \emph{synphot tool} of the \emph{STSDAS} software package\footnote{\url{http://www.stsci.edu/institute/software_hardware/stsdas}}. Reported spectral slopes are the mean values from all of a target's measured values, weighted by the inverse of the uncertainty in each measurement. In all cases that an object has several independent colour measurements, all measurements were in agreement to within $2-\sigma$ of the quoted uncertainties. If an individual measurement was more precise than the weighted mean, that value was used instead. We consider only those measurements with spectral slope uncertainty $\Delta s < 7\%$, the largest colour uncertainty for any of the known binary CCKBOs. The full list of colour measurements we consider is presented in Supplementary Table 1.

Measurements of 11 new CCKBOs were gathered with the GMOS detector\cite{AS_2002,Hook_2004} on the Gemini-North telescope as part of the program Colours of the Outer Solar System Origins Survey (Col-OSSOS). The sample consists of all CCKBO targets with brightness at discovery $r'<23.6$ as observed by the Outer Solar System Origins Survey\cite{Bannister_2016}. Orbit classification of these targets was done as above, confirming their non-resonant behaviour. Photometry was acquired in the Sloan g' and r' filters in a rg and a gr sequence using 300~s exposure times, with total number of exposures tuned to achieve a final photometric precision in the (g'-r') of no more than 0.06 magnitudes, or $\Delta s \lesssim 3\%$. Images were preprocessed with standard techniques and photometrically calibrated using background stars catalogued in the Sloan Digital Sky Survey (SDSS)\cite{Alam_2015}. Photometry was measured using the \emph{TRIPPy} package \citeM{Fraser_2016}. To account for small lightcurve variations across the observing sequences, a line was fit to the g and r measurements simultaneously, assuming a constant (g'-r') colour, using standard least-squares techniques (see Supplementary Figure 1). The mean fit colour was then converted to the SDSS system using a linear colour conversion between the Sloan and Gemini filters. This conversion was determined to be $(g'-r')_{\textrm{Gemini}}=0.90(\pm0.01)\times (g'-r')_{\textrm{SDSS}}$. \emph{TRIPPy} was used to calculate model trailed point sources for each target. The brightness of each source was fit in a maximum likelihood sense to the actual image and removed. The residual images were visually inspected for candidate binaries. For objects 2016 BP81, 2013 SQ99, and 2014 UD255, the secondary was obvious (see Fig. 1). The fitting procedure was repeated using two model trailed sources simultaneously to measure binary component separation and brightness ratio. No secondaries were detected around other sources to the noise limit of the data. Primary-secondary brightness ratios are presented in Supplementary Table 2.

Images of 2002 VD131, the bluest known CCKBO not previously identified as a binary \citeM{Gulbis_2006}  were acquired with MegaCam\cite{Boulade_2003} on the Canada-France-Hawaii Telescope (CFHT) on January 1, 2016, and with IMACS\cite{Dressler_2006} on the Baade-Magellan Telescope on December 15, 2015. Photometry was measured and binarity was searched for, as above. VD131 exhibits a faint companion with relative position and brightness that are consistent in both datasets. The CFHT and Magellan imagery resulted in $5$ and $8-\sigma$ detections of the secondary, respectively. 

Colours of binaries 2003 UN284, 2005 EO304, 2006 BR284, 2006 CH69, and 2006 JZ81 were measured from photometry acquired for binary orbit determination \citeM{Parker_2011} using the techniques discussed above. No conversion between the Very Large Telescope and SDSS filter systems is known, so we restrict photometric calibration to SDSS stars with $0.5<(g'-r')<1.0$, similar to most KBOs. Where insufficient background stars were available, zeropoints from other images taken on the same night were used, and an additional uncertainty of 0.03 magnitudes was included, accordingly. All spectral slopes are reported in Supplementary Table~1.

Where available, binary semi-major axes were taken from the \emph{Mutual Orbits of Trans-Neptunian Binaries} webpage\footnote{\url{http://www2.lowell.edu/~grundy/tnbs/}} \citeM{Grundy_2011}. Where the orbit has not yet been reliably determined, the $3-\sigma$ lower-limit to the binary semi-major axis was taken as $a_{\textrm{lim}}=\frac{r}{1+e_{\textrm{max}}}$ where $r$ is the $3-\sigma$ lower limit of the widest observed separation of a target. We adopt $e_{\textrm{max}}=0.8$. All semi-major axes and lower-limits are reported in Supplementary Table~2.

\subsection{Colour Distribution Comparison}

We address the question of the null hypothesis that the observed binary and single colour distributions are the same. From the observed colour distribution, it is apparent that the binary colour distribution is wider, has a bluer mean colour, and extends to bluer values than does the single colour distribution. Thus, between the binary sample of 29, and the single sample of 58, we utilize the absolute difference of means and the difference of widths statistics.  To that end, from the distribution of 58 singles colour sample, we bootstrapped with repetition samples of 29 objects. The absolute difference of the means and standard deviations of the random sample and the observed single population were recorded. The difference of means statistic suggests there is only a 0.09\% chance that the binary and single objects share the same colour distribution. The difference of standard deviations suggests that the probability is 0.1\%. 

We also define the observed condition, that the sample of 29 has 5 or more objects that are bluer than the bluest object in the sample of 58. When bootstrapping two random samples of 29 and 58 from the full sample, the observed condition occurred in only 0.11\% of those simulations.

In our testing, we avoided the standard Kolmogorov-Smirnov and Anderson-Darling tests which are insensitive to distributions that differ maximally in their tails. We also caution that our tests implicitly assume that the binary and single colour distributions are biased in the same manner. The biases in our colour distributions are hard to quantify. Typically, once a target has been identified as a binary, it receives additional focused observations for various purposes, which may increase the chance of it receiving a colour measurement compared to an otherwise similar, but single CCKBO. 

\subsection{Binary Migration Simulations}
It is now well accepted that the majority of the dynamical structure of the Kuiper Belt was acquired during the early migration of the gas-giant planets, and in particular, that of Neptune (eg.\citeM{Levison_2003}). In terms of matching the known orbital structure of the CCKBOs, the simulations of \citeM{Nesvorn__2015} are the most successful. Those simulations are characterized by an initial period of fast migration, and a later period of slower migration onto its current orbit, with a small jump of $\sim0.5$~AU in semi-major axis and 0.05 in eccentricity which occurs when Neptune reached 27.8 AU. During the initial migratory phase, planetesimals in the 35-40~AU range are temporarily swept into the 2:1 MMR, and transported outwards. When the jump occurs, some objects leave the 2:1 and are deposited into the CCKBO region. The final stage of migration and further evolution to the current age of the Solar System causes minor excitation to objects in that region.

We probe if binary KBOs can survive the migratory scenario presented by \citeM{Nesvorn__2012} by repeating the nominal simulation presented in \citeM{Nesvorn__2015}. We made use of the leapfrog integration scheme implemented in the REBOUND integrator\footnote{\url{https://github.com/hannorein/rebound}}. We implemented the fictitious damped migratory forces of \citeM{Wolff_2012}, which on secular timescales produce semi-major axis, eccentricity, and inclinations given by

\begin{align*}
e&=e_o\exp{\frac{-t}{\tau_e}},\\
i&=i_o\exp{\frac{-t}{\tau_i}},\\
a&=a_f+\Delta a\exp{\frac{-t}{\tau_a}},
\end{align*}

\noindent where $e_o$, $i_o$, $a_f$ are the initial eccentricity, inclination, and final semi-major axes of the forced body. $\Delta a$ sets the approximate total distance the body would migrate during the integration. We also include an additional gravitational force term between pairs of initially bound planetesimals. The integrations utilized a 0.5 day timestep to ensure that the binary orbits with orbital periods as short as $\sim11$~days were evolved correctly.

Guided by \citeM{Nesvorn__2015} and \citeM{Nesvorn__2016}, for the initial phase of migration we adopt $\tau_e=\tau_i=\tau_a=\tau=30$~Myr, $i_o=1^\circ$, $e_o=0.01$, $\Delta a=7$~AU, and $a_f=30.07$~AU. At 39.12 Myr when Neptune first reaches 27.8 AU, the nominal location of the jump, we halt the simulation.

Our simulations were tuned to test how well binaries with a range of separations can survive the migration of Neptune. Simulations started with 125 equal mass binary pairs with total system mass of $10^{18}$~kg typical of the widest separated CCKBO binaries, like 2001 QW322. Initial binary semi-major axes were randomly selected uniformly through the current observed range. Three sets of simulations were run. Set 1 consisted of 18 iterations with moderate binary semi-major axes sampled from the range $1,500\leq a_{\textrm{bin}} \leq40,000$~km. Set 2 consisted of 6 iterations of 125 binaries with wide initial semi-major axes, $40,000\leq a_{\textrm{bin}} \leq100,000$~km. Set 3 consisted of 9 iterations of 125 binaries with $1,500\leq a_{\textrm{bin}} \leq15,000$~km to improve the survival rate measurement for the initially closest separated binaries. Binaries were placed on circular orbits with random orientations. Heliocentric inclinations and eccentricities were randomly drawn from the Brown distribution $\sin{i} \exp{-\frac{i^2}{2\sigma_i^2}}$ and a Rayleigh distribution respectively, with $\sigma_i=2^\circ$ and Rayleigh parameter $\sigma_e=0.01$.

We show the resultant eccentricity distribution and survival rate of planetesimals that ended in the cold classical region in Supplementary Figure. 2. The majority of objects from both simulations experienced close encounters with Neptune, and were either ejected, or implanted on dynamically excited orbits. The overall population had a survival rate of $\sim0.5$\%. No binaries with $a_{\textrm{bin}} \gtrsim0.25 \mbox{ $R_{\textrm{H}}$}$ however, survived. The surviving binaries had initial heliocentric semi-major axes as low as $a=38$~AU, with no obvious correlation between initial and final heliocentric orbital elements. Most binaries experienced a substantial excitation in binary eccentricity, with final values reaching as high as $0.8$; only those objects with $a_{\textrm{bin}} \lesssim 0.1 \mbox{ $R_{\textrm{H}}$}$ experienced virtually no eccentricity excitation.

We chose to forgo the second stage of migration because the excitation of the cold classical objects is sensitive to Neptune's unknown, and not very well constrained late migratory behaviour \citeM{Batygin_2011}. Simulations by Nesvorny however, have demonstrated that regardless of the behavior, only modest change in the orbits of the cold classical objects occurs. In his nominal simulations, the mean change in eccentricity and inclination of objects that were in the cold classical region at the start and end of the second stage of migration was -0.05 and $-0.3^\circ$ with a survival rate of $\sim65\%$. Thus, while the exact orbital distribution cannot be extracted from ours or other migration simulations, the lack of simulation of the second stage of migration does not alter our conclusion that widely separated binaries can survive push-out by a migrating Neptune.

To see if the combined effects of collisional and tidal evolution of binary systems could account for the lack of closely bound blue binaries, we analyzed the results of \citeM{Brunini_2015} in search of any trends of change in binary semi-major axis with initial binary eccentricity. As can be seen in Supplementary Figure. 3, no significant mean change in binary semi-major axis occurred for systems starting with eccentricities $\lesssim 0.7$. It appears that the combined effects of collisional and tidal evolution together, alongside the eccentricity excitation experienced during migration, cannot account for the lack of observed closely separated blue binaries.

\subsection{Binary Formation and Cloud Collapse Simulations}

Various KBO binary formation mechanisms have been put forth to remove angular momentum and bind two single bodies in close proximity. These mechanisms invoke scattering of one or more smaller bodies \citeM{Goldreich_2002,Astakhov_2005}, or collisions between two KBOs within the sphere of influence of a third \citeM{Weidenschilling_2002} to bind two of those objects into a binary system. Given the poor efficiency of some of these mechanisms \citeM{Schlichting_2008} and the necessity of a third body, it seems implausible that the conditions were such that most of these mechanisms could successfully convert virtually all single bodies into binaries \citeM{Nesvorn__2010}. Nesvorny et al (2010) point out that to achieve binary formation rates sufficient to account for the CCKBO binary fraction, the planetesimal velocity dispersion must remain sub-hill during planetesimal growth. These conditions are met by the pebble accretion model of \citeM{Shannon_2016}. When the velocity dispersion and pebble densities of that model are considered, the L2s formation mechanism predicts high fractional binary conversion rates of $\sim10^{-6}$ per year. While it is not clear if the L2s model can ever achieve high enough of a binary fraction to be compatible with the $\sim100\%$ binary fraction exhibited by the blue binaries, from rate estimates alone, the L2s mechanism remains a viable formation route.

An alternate formation route is presented by gravitational collapse. \citeM{Nesvorn__2010} simulate the formation of large planetesimals during the collapse of gravitationally bound clouds of small objects. This process preferentially produces bound high multiplicity systems, offering a viable alternative mechanism for binary production. We analyzed the collisional behaviours of those simulations under varied initial conditions. Three separate simulations were run with initial mass of $14.1 \times 10^{18}$~kg (assuming density of $1 \mbox{ g cm$^{-3}$}$) and three separate values of initial angular momentum. The outcomes of those simulations are presented in Supplementary Table~3. In Supplementary Figure. 4 we present one example of the accretionary collisions that drove the growth of a triple system that resulted from one such simulation. Assuming a density of $\rho=1 \mbox{ g cm$^{-3}$}$, at the end of the simulation, the combined mass of the two largest bodies is $M_{\textrm{tot}} = 5.5\times10^{18}$~kg, with a binary semi-major axis $a_{\textrm{bin}}=5280$~km,  similar to 2000 QL251 (mass and binary semi-major axis of $3\times10^{18}$~kg and 5000~km). 

In all three simulations, the radii of the largest objects, their total system mass, and the impact velocities experienced by those large bodies were similar: radii, maximum impact velocities, and system masses were within factors of 1.5, 2, and 3 of each other (assuming equal densities). Notable changes were found however, in overall formation efficiency, and final binary semi-major axes. For these three simulations, an increase in angular momentum results in an increase in the fraction of initial cloud material that is ejected, and conversely, a decrease in the fraction of initial cloud mass accreted into the large remaining bodies. An increase in angular momentum also results in increased semi-major axis of the two largest components. This trend suggests that the orbital differences seen between the blue and red CCKBO binary systems may be a result of the conditions at the start of the collapse, the former starting with typically large cloud masses with higher initial angular momenta than the latter.

Gravitational collapse of a pebble cloud occurs when the Roche density is reached inside the cloud. At that point, the cloud has mass $M_{\textrm{tot}}$, with intial radius $\sim R_\textrm{H}$ and density $\rho_\textrm{R}$. $R_\textrm{H}$ is the Hill radius, which is proportional to the heliocentric distance and the cube root of cloud mass; the Roche density, $\rho_\textrm{R}$, is proportional to the inverse heliocentric distance cubed.

\begin{displaymath}
R_\textrm{H} \propto a \, M_{\textrm{tot}}^{1/3}, \quad \rho_\textrm{R} = \frac{9\Omega^2}{4\pi G} \propto \left(\frac{a}{\textrm{AU}}\right)^{-3}
\end{displaymath}

The critical spin frequency above which the pebble cloud will disperse scales as the square root of its density. 
\begin{displaymath}
\omega_\textrm{crit} \propto \rho^{1/2} = \rho_\textrm{R}^{1/2} \propto a^{-3/2}
\end{displaymath}

At that point, the cloud angular momentum is
\begin{align*}
L_\textrm{crit} &\propto M_{\textrm{tot}}\,r^2\,\omega \\
&= m \, R_\textrm{H}^2 \,\omega_\textrm{crit} \\
&= m \, a^2\,M_{\textrm{tot}}^{2/3} \, a^{-3/2} \\
&= M_{\textrm{tot}}^{5/3} \, a^{1/2}
\end{align*}

For the angular momentum of pebble clouds to decrease with heliocentric distance, the cloud mass must also decrease with distance, faster than $a^{-3/10}$.

\subsection{Ejecta Transfer and Binary Disruption Simulations}

Simulations of ejecta transfer within a binary system were performed. Particles were ejected with a range of velocities from one body, and integrated under the influence of the gravity of both binary bodies, the Sun, and radiation pressure. Ejection velocities and radiation pressure parameters $\beta$ were chosen at random between 1-3 times the escape velocity of the impacted body, and 0 to 1, respectively. Ejected particles would arrive at the other body over a period of many tens of hours. For all values of $\beta$, the probability of impact was equal to the fraction of area subtended by the second body projected to the binary separation distance. 

The velocity and ejecta distribution scalings presented in \citeM{Leinhardt_2012} were used to estimate the mass ejected from the impacted body. We consider collisions onto a system like 2001 QW322 - the most widely separated blue binary - with roughly equal sized components, and separation of $\sim10^6$~km. For a 16\% albedo typical of CCKBOs \citeM{Fraser_2014} the components of the QW322 system have radii $r\sim35$~km. 

Considering a $v_{\textrm{imp}}= 500 \mbox{ m s$^{-1}$}$ collision onto one of the components, an impactor with radius $\sim4$~km is required to uniformly coat the non-collided binary member in a $3 \mbox{ $\mu$m}$ thick layer, the approximate depth of the optically active surface. Roughly, 100 collisions by $1$~km impactors are required to achieve the same depth of transferred material. For even the highest collision velocities of $\sim30 \mbox{ m s$^{-1}$}$, there is virtually no mass transfer between the components of the widest separated binaries. For both cases however, multiple orders of magnitude more ejected material coats the impacted target compared to the non-impacted component; for $v_{\textrm{imp}} = 500 \mbox{ m s$^{-1}$}$, an impact by an $\sim 0.1$~km impactor is required to recoat the entire surface of the impacted body.

The observed KBO size distribution \citeM{Fraser_2014} has a rapidly increasing number of objects with decreasing size. Thus, collisions that would resurface only the more readily coated primary will occur more often than collisions that would resurface both objects. This is true for both scenarios in which the impactors originate externally, or are gravitationally bound to the system.

\end{methods}


\newpage

\textbf{Methods References}


\begin{figure}
\includegraphics[width=15cm]{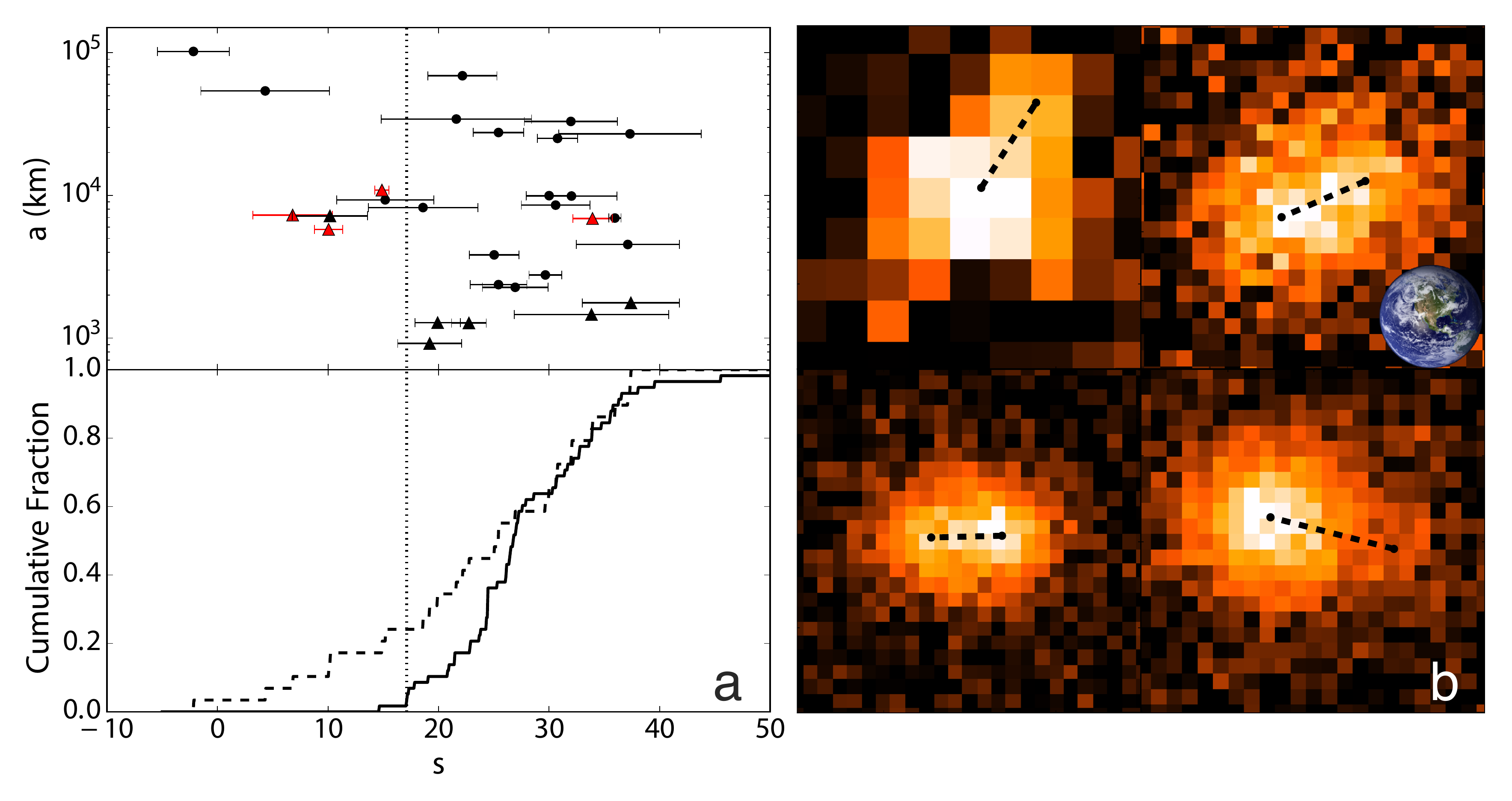}
\caption{\textbf{Left} Top: Binary semi-major axis versus optical spectral slope, $s$, of known CCKBO binary objects with well determined colours. We quantify a target's colour with spectral slope, $s$, defined as percent increase in reflectance per 100~nm change in wavelength normalized to 550~nm. Points in red are new binaries presented here. Round points indicate systems for which the binary semi-major axis has been determined. Triangles are lower limits on semi-major axis (see Methods). Bottom: Cumulative spectral slope distribution of single (58 objects, solid line) and binary cold classical objects (29 objects, dashed line). The vertical dotted line is the spectral slope that divides the blue and red classes of the dynamically excited KBOs. \textbf{Right} Images of the four new binaries, scaled to the same relative distance scale. Black lines show the fitted distances of the two components. The points are roughly 5x larger than the true sizes of the objects. Clockwise from top-left, 2002 VD131, 2016 BP81, 2014 UD255, and 2013 SQ99. The Earth, with mean diameter 12,742~km is shown for scale.}
\end{figure}

\begin{figure}
\includegraphics[width=15cm]{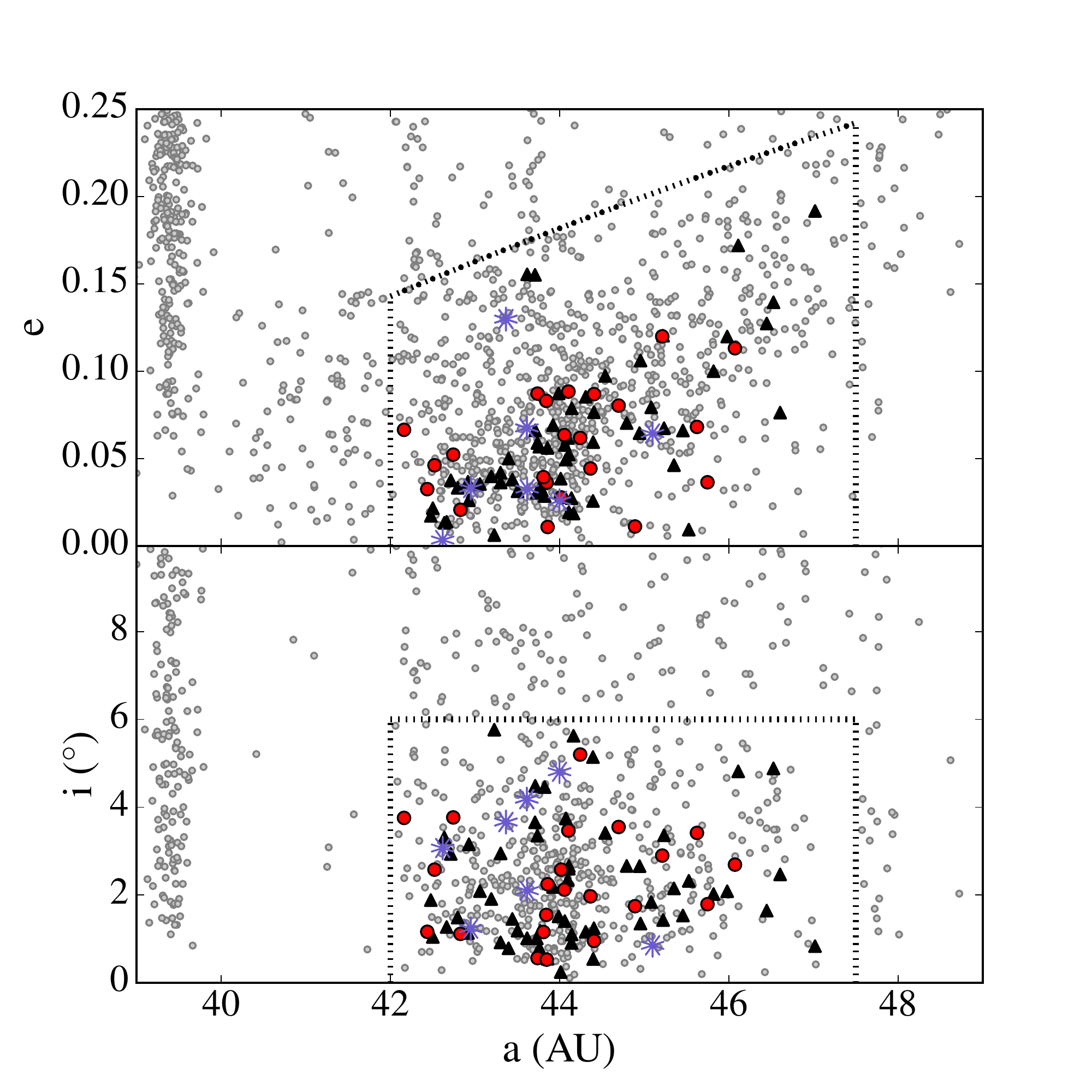}
\caption{Barycentric orbital elements, eccentricity (top) and inclination (bottom) vs. semi-major axis of KBOs. The dashed lines show the boundaries of the cold classical region we adopt (see Methods). Coloured points are CCKBOs with well measured colours. Grey points are objects with no reliable colour measurement. Black triangles, red circles, and blue stars represent single, red binary  ($s>17\%$), and blue binary ($s<17\%$) CCKBOs.}
\end{figure}

\begin{figure}
\includegraphics[width=15cm]{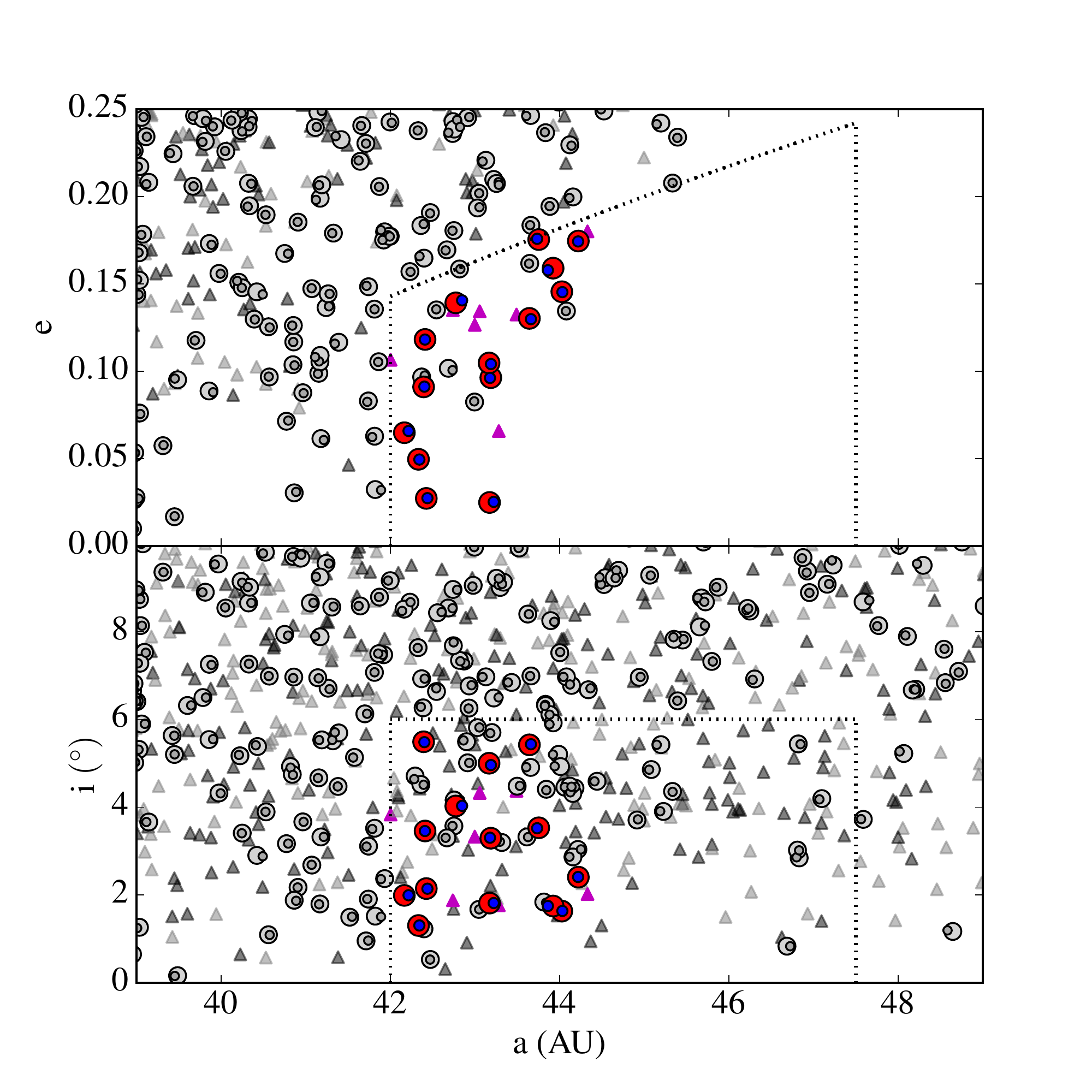}
\caption{Barycentric orbital elements of the surviving particles immediately after Neptune's jump, at 27.8~AU. Dotted lines demark the cold classical region. Pairs of overlapping large and small round points mark bound binary pairs and triangles mark single objects all of which are the result of binary unbinding. Red-blue pairs and purple triangles are those binary and single objects which were emplanted in the cold classical region. As in \cite{Levison_2003}, some objects transported outward into the cold classical region fell out of the 2:1 MMR before the jump due to Neptune's non-smooth migration, while others dropped out of the resonance when the planet jumped.}
\end{figure}

\spacing{1}

\renewcommand{\tablename}{Supplementary Table}
\renewcommand{\thetable}{\arabic{table}}
\renewcommand{\figurename}{Supplementary Figure}
\renewcommand{\thefigure}{\arabic{figure}}
\setcounter{figure}{0}

\begin{figure}[t]
\includegraphics[width=15cm]{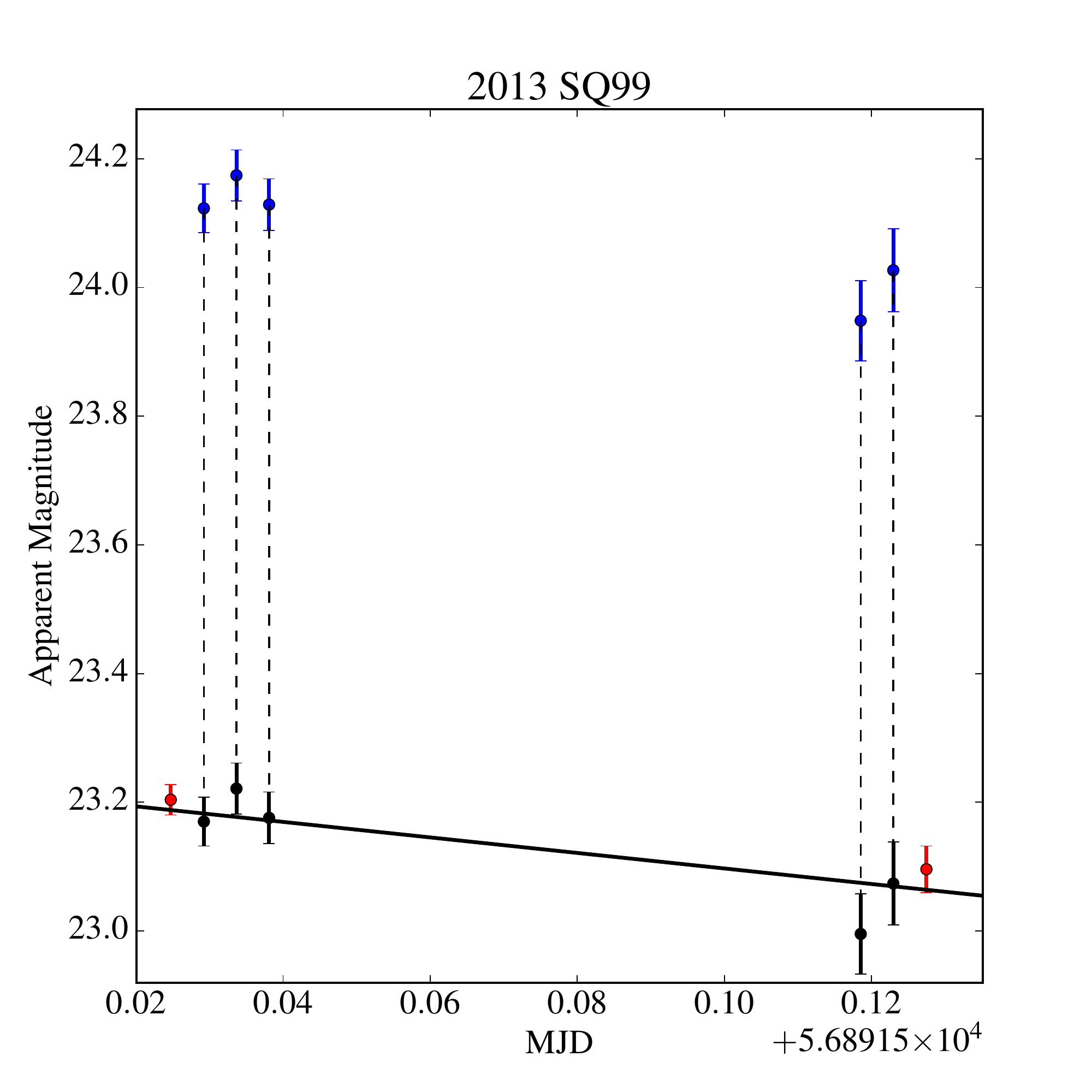}
\caption{Observed apparent magnitudes of binary 2013 SQ99 in the r' (red) and g' (blue) filters. Black points show r' magnitudes estimated from the g' observations and the mean (g'-r')$=0.89\pm0.02$ colour in the Gemini filters. The best-fit linear lightcurve which decreases $\sim0.04$ mags from start to end of sequence is shown by the solid black line.}
\end{figure}

\begin{figure}[h]
\includegraphics[width=15cm]{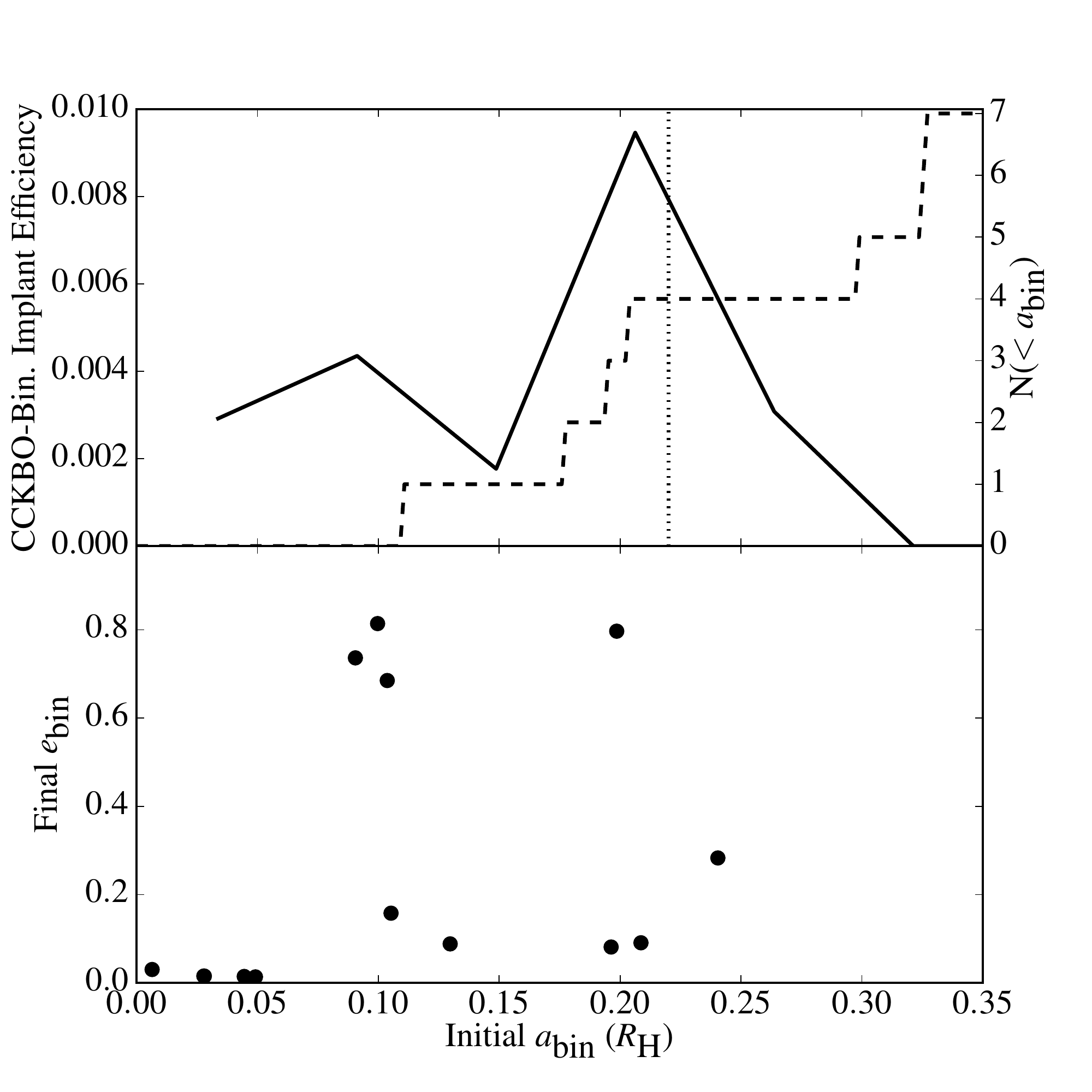}
\caption{Top: the solid line presents the efficiency of implantation by migration of binaries into the cold classical region as a function of initial binary semi-major axis. The dashed line presents the cumulative distribution of singles implanted in the cold classical region as a function of initial binary semi-major axis. The vertical dashed line presents the current binary semi-major axis of the widest separated known, cold classical binary, 2001 QW322. Bottom: final eccentricity of CCKBO binary survivors vs. initial binary semi-major axis.}
\end{figure}

\newpage

\begin{figure}[h]
\includegraphics[width=15cm]{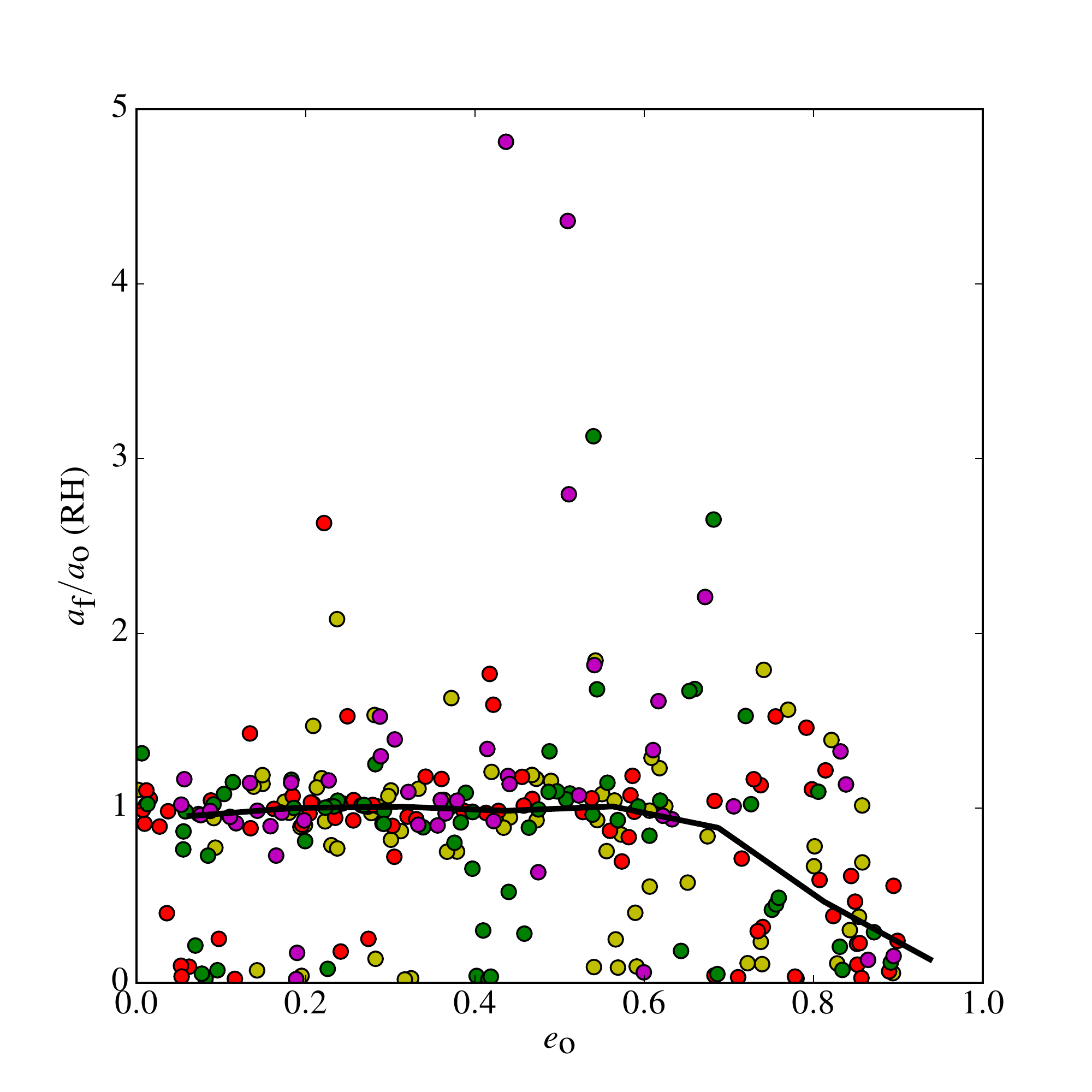}
\caption{Ratio of final and initial binary semi-major axis vs. initial binary eccentricity for the binary evolution simulations of Bruini and Zanardi (2016) which assumed a tidal quality factor of Q=10. Yellow, red, green, and magenta points correspond to binary size ratios of 0.25, 0.5, 0.75, and 1. The black line depicts the mean ratio as a function of initial eccentricity for the entire sample. A mean reduction in semi-major axis is observed for eccentricities  $e_\textrm{o}\gtrsim 0.6$ where tidal forces become the dominant driver of the binary evolution.}
\end{figure}

\newpage

\begin{figure}[h]
\includegraphics[width=15cm]{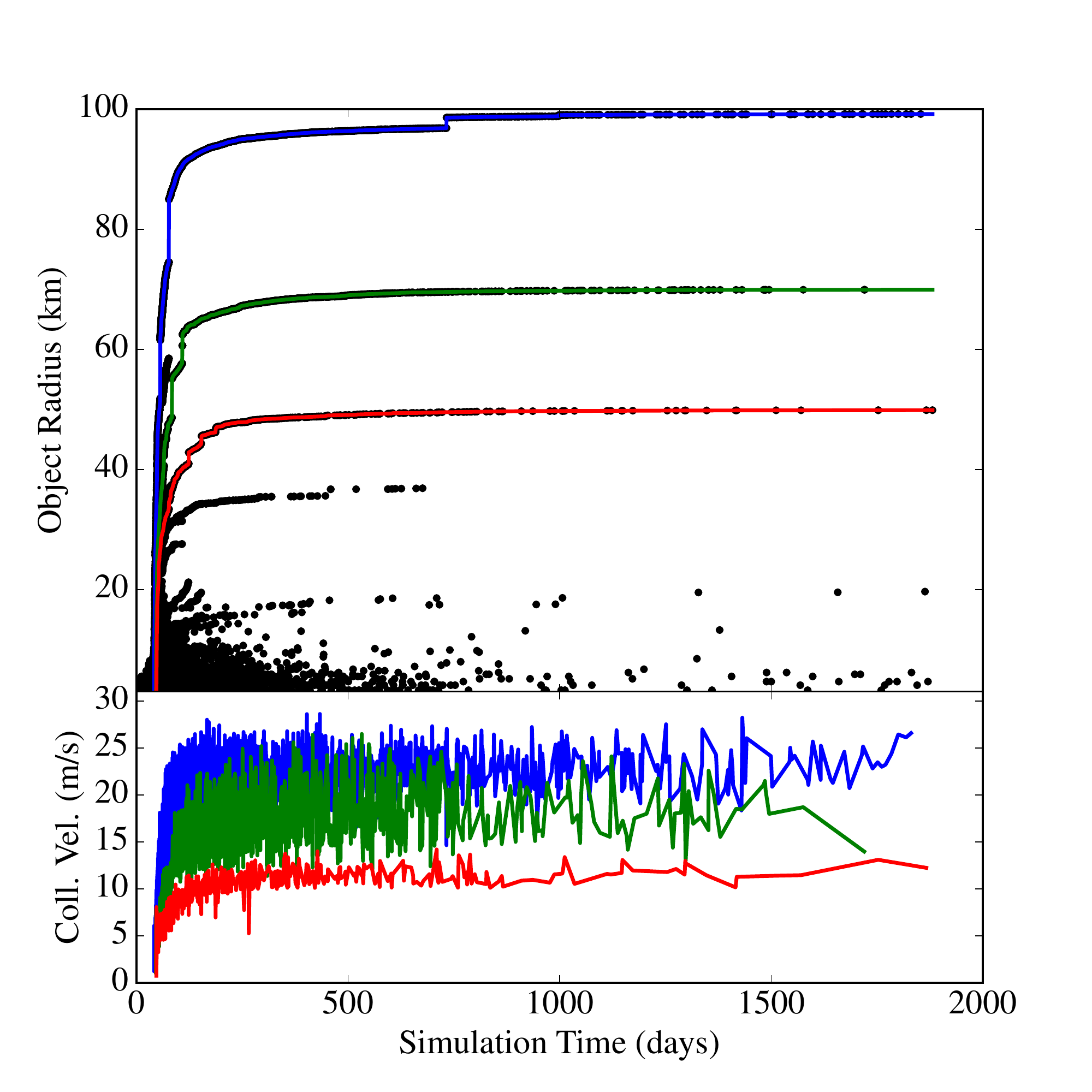}
\caption{The collisional growth during the collapse of a gravitationally bound cloud. The growth and collisions experienced by the largest three bodies which the simulation produces are traced by the blue, green, and red lines, respectively. The top panel displays the evolution of the radius of each object as a function of time. Black dots denote when a collision occurs, and the radius of the merged body. The bottom panel displays the collision velocities of each collision in the top panel. The collision velocities are very similar regardless of impactor size. As exhibited by the nearly smoothly increasing radius of the large bodies with time, impacts are dominated by the smallest bodies in the simulation.}
\end{figure}


\textbf{Supplementary Information for: All planetesimals born near the Kuiper Belt formed as binaries}\\

\begin{minipage}{\linewidth}
\centering
\scriptsize
\captionof{table}{Cold Classical KBO Optical Spectral Slopes}
\begin{tabular}{l c l c }
\toprule[1.5pt]
 \bf Target & \bf $s$ &  \bf Target & \bf $s$ \\ \midrule
 \multicolumn{4}{c}{Single Objects}\\ \midrule
15760 - 1992 QB1 & $23.8 \pm 2.3$ & 2000 FS53 & $33.9 \pm 2.0$ \\
16684 - 1994 JQ1 & $35.8 \pm 3.0$ & 2000 OH67 & $24.5 \pm 5.1$ \\
19255 - 1994 VK8 & $26.1 \pm 3.1$ & 138537 - 2000 OK67 & $19.0 \pm 1.9$ \\
1995 DC2 & $45.5 \pm 1.4$ & 2000 QC226 & $51.3 \pm 5.6$ \\
1995 WY2 & $24.3 \pm 6.0$ & 2001 HZ58 & $22.8 \pm 5.4$ \\
1996 TK66 & $26.7 \pm 3.9$ & 88268 - 2001 KK76 & $27.1 \pm 3.6$ \\
1997 CT29 & $38.1 \pm 5.1$ & 2001 OQ108 & $30.3 \pm 4.4$ \\ 
33001 - 1997 CU29 & $33.8 \pm 2.1$ & 2001 QE298 & $33.6 \pm 2.3$ \\
52747 - 1998 HM151 & $27.9 \pm 6.0$ & 2001 QO297 & $32.2 \pm 3.7$ \\
85627 - 1998 HP151 & $26.5 \pm 6.3$ & 2001 QP297 & $27.6 \pm 1.2$ \\
385194 - 1998 KG62 & $31.7 \pm 3.6$ & 2001 QR297 & $24.5 \pm 4.0$ \\
85633 - 1998 KR65 & $30.7 \pm 1.5$ & 2001 QS322 & $22.9 \pm 3.3$ \\
1998 KS65 & $27.2 \pm 2.4$ & 2001 QX297 & $24.5 \pm 3.5$ \\
69987 - 1998 WA25 & $17.1 \pm 6.0$ & 126719 - 2002 CC249 & $20.8 \pm 4.6$ \\
1998 WX24 & $35.5 \pm 4.3$ & 2002 CU154 & $24.5 \pm 4.0$ \\
1998 WY24 & $26.2 \pm 2.6$ & 2002 PD155 & $17.8 \pm 5.6$ \\
1999 HG12 & $26.4 \pm 1.7$ & 2002 PV170 & $23.7 \pm 2.2$ \\
1999 HS11 & $32.8 \pm 3.3$ & 385437 - 2003 GH55 & $24.5 \pm 1.8$ \\
1999 HV11 & $24.3 \pm 2.3$ & 385447 - 2003 QF113 & $25.3 \pm 3.4$ \\
385199 - 1999 OE4 & $14.6 \pm 5.0$ & 2003 QY111 & $17.3 \pm 5.6$ \\
66452 - 1999 OF4 & $30.6 \pm 6.0$ & 2006 HW122 & $17.1 \pm 6.1$ \\
455171 - 1999 OM4 & $21.5 \pm 3.3$ & 2006 QF181* & $26.4 \pm 1.2$ \\
1999 RC215 & $39.5 \pm 5.3$ & 2013 SP99* & $34.7 \pm 0.6$ \\
137294 - 1999 RE215 & $36.3 \pm 2.5$ & 2013 UL15* & $26.1 \pm 1.8$ \\
1999 RX214 & $20.9 \pm 1.0$ & 2013 UM15* & $32.7 \pm 1.8$ \\
2000 CE105 & $28.6 \pm 4.3$ & 2013 UN15* & $36.6 \pm 2.4$ \\
60454 - 2000 CH105 & $26.8 \pm 1.2$ & 2013 UO15* & $27.0 \pm 1.2$ \\
2000 CL104 & $21.4 \pm 2.4$ &  2013 UP15* & $27.2 \pm 0.6$ \\
2000 CN105 & $31.4 \pm 3.8$ & 2014 UE225* & $35.6 \pm 0.6$ \\

\multicolumn{4}{c}{Binary Objects} \\ \midrule
58534 - 1997 CQ29 & $18.6 \pm 5.0$ & 148780 - 2001 UQ18 & $32.0 \pm 4.1$ \\
79360 - 1997 CS29 & $29.7 \pm 1.5$ & 2002 VD131 & $6.8 \pm 3.6$ \\
1999 RT214 & $37.4 \pm 4.4$ & 2002 VT130 & $22.7 \pm 1.6$ \\
66652 - 1999 RZ253 & $37.1 \pm 4.7$ & 2003 HG57 & $10.2 \pm 3.4$ \\
2000 CF105 & $21.6 \pm 6.8$ & 2003 QY90 & $30.6 \pm 3.1$ \\
80806 - 2000 CM105 & $33.8 \pm 7.0$ & 2003 TJ58 & $25.0 \pm 2.2$ \\
2000 CQ114 & $36.0 \pm 0.6$ & 2003 UN284 & $4.3 \pm 5.8$ \\
134860 - 2000 OJ67 & $26.9 \pm 3.0$ & 2005 EO304 & $22.2 \pm 3.1$ \\
123509 - 2000 WK183 & $25.4 \pm 2.6$ & 2006 BR284 & $30.8 \pm 1.8$ \\
2000 WT169 & $19.9 \pm 2.0$ & 2006 CH69 & $37.3 \pm 6.5$ \\
88611 - 2001 QT297 & $25.4 \pm 2.3$ & 364171 - 2006 JZ81 & $32.0 \pm 4.2$ \\
2001 QW322 & $-2.2 \pm 3.3$ & 2013 SQ99* & $33.9 \pm 1.8$ \\
275809 - 2001 QY297 & $30.0 \pm 2.1$ & 2014 UD225* & $14.9 \pm 0.6$ \\
2001 RZ143 & $19.2 \pm 2.9$ & 2016 BP81* & $10.1 \pm 1.3$ \\
 2001 XR254 & $15.2 \pm 4.4$ & \\

\bottomrule[1.25pt]
\end{tabular}\par
\bigskip
Note: Objects that are classified as blue have $s<17\%$. Those objects with non-MPC designations are internal OSSOS designations of cold classical objects. * - colours measured by Col-OSSOS.
\end{minipage}

\begin{minipage}{\linewidth}
\centering
\footnotesize
\captionof{table}{Observed Binary Properties}
\begin{tabular}{l c c}
\toprule[1.5pt]
 \bf Target & \bf Separation (",km) & \bf Secondary:Primary Brightness Ratio \\ \midrule
2002 VD131 & $0.48\pm0.02$, $14900\pm600$ & $0.061^{+0.012}_{-0.004}$\\
2016 BP81 & $0.369^{+0.02}_{-0.01}$,$11300^{+600}_{-300}$ & $0.81^{+0.09}_{-0.06}$ \\
2013 SQ99 & $0.39^{+0.02}_{-0.01}$, $13300^{+700}_{-300}$ & $0.65^{+0.06}_{-0.05}$\\
2014 UD225 & $0.67\pm0.02$, $21400\pm600$ & $0.12\pm0.01$\\
\bottomrule[1.25pt]
\end{tabular}\par
\bigskip
Note: Uncertainties are $1-\sigma$ uncertainties. The confidence intervals have been combined using all images available for each target for the quoted uncertainties. Binary properties were measured from Gemini-GMOS images of 2013 SQ99 and 04h45. For 2016 BP81, only the single GMOS frame with the best image quality was used as it was the only frame to resolve the two components. The single images with best image quality from each of the Magellan and CFHT image sequences were used to measure the properties of 2002 VD131.
\end{minipage} 

\vspace{2 cm}

\begin{minipage}{\linewidth}
\centering
\footnotesize
\captionof{table}{Initial and final properties of example cloud collapse simulations}
\begin{tabular}{c c c c}
\toprule[1.5pt]
\bf $\Omega \mbox{ $(\Omega_{\textrm{circ}})$}$ & $r_{\textrm{p}}$ (km) & $r_{\textrm{s}}$ (km) & $a_{\textrm{bin}}$ (km) \\ \midrule
 0.1 & 138.5 & 66.8 & 4731\\
 0.5 & 99.2 & 69.9 & 5281\\
 0.75 & 90.2 & 65.9 & 7066\\
\bottomrule[1.25pt]
\end{tabular}\par
\bigskip
Note: Initial mass $M_{\textrm{tot}}$ and cloud angular velocity are presented. $\Omega_{\textrm{circ}}=\sqrt{\frac{G M_{\textrm{tot}}}{R_{\textrm{tot}}^3}}$ where $R_{\textrm{tot}}$ is the initial cloud radius. Final primary and secondary radii and binary semi-major axis at end of the simulations are also presented.
\end{minipage}\\

\end{document}